\documentclass[twocolumn,showpacs,preprintnumbers,amsmath,amssymb,prb,superscriptaddress]{revtex4}
\usepackage[T1]{fontenc}
\usepackage[latin1]{inputenc}
\usepackage{graphicx,color}

\newcommand{\buck}{{C$_{60}$}}

\newcommand{\beq}{\begin{equation}}
\newcommand{\eeq}{\end{equation}}

\newcommand{\beqa}{\begin{eqnarray}}
\newcommand{\eeqa}{\end{eqnarray}}

\DeclareMathOperator{\tr}{Tr}
\newcommand{\etal}{\mbox{\textit{et al.}}}

\newcommand{\Eqref}[1]{Eq.~(\ref{#1})}
\newcommand{\Figref}[1]{Fig.~\ref{#1}}
\newcommand{\Secref}[1]{Sec.~\ref{#1}}

\makeatother

\begin{document}

\title{Nonequilibrium electron-vibration coupling and conductance fluctuations\\
in a \buck-junction}
\author{S\o ren~Ulstrup}\altaffiliation{Present address: Department of Physics and Astronomy, Interdisciplinary
Nanoscience Center, Aarhus University,
8000 Aarhus C, Denmark}
\affiliation{DTU-Nanotech, Department of Micro- and Nanotechnology, Technical
University of Denmark, {\O}rsteds Plads, Bldg.~345E, DK-2800 Kongens
Lyngby, Denmark}
\author{Thomas~Frederiksen}
\affiliation{Donostia International Physics Center (DIPC) -- UPV/EHU, E-20018 Donostia-San
Sebasti\'an, Spain}
\affiliation{IKERBASQUE, Basque Foundation for Science, E-48011, Bilbao, Spain}
\author{Mads~Brandbyge}
\affiliation{DTU-Nanotech, Department of Micro- and Nanotechnology, Technical
University of Denmark, {\O}rsteds Plads, Bldg.~345E, DK-2800 Kongens
Lyngby, Denmark}
\email{mads.brandbyge@nanotech.dtu.dk}
\date{\today}

\begin{abstract}
We investigate chemical bond formation and conductance in a molecular
\buck-junction under finite bias voltage using first-principles calculations based
on density functional theory and nonequilibrium Green's functions (DFT-NEGF).
At the point of contact formation we identify a remarkably strong coupling between
the \buck-motion and the molecular electronic structure. This is only seen for
positive sample bias, although the conductance itself is not strongly polarity
dependent. The nonequilibrium effect is traced back a
sudden shift in the position of the voltage drop with a small \buck-displacement.
Combined with a vibrational heating mechanism we construct a model from our results that explain the
polarity-dependent two-level conductance fluctuations observed in recent scanning
tunneling microscopy (STM) experiments [N.~N\'eel \etal, Nano Lett.~{\bf 11},
3593 (2011)].
These findings highlight the significance of nonequilibrium effects in chemical bond
formation/breaking and in electron-vibration coupling in molecular electronics.
\end{abstract}

\maketitle
\section{Introduction}

The influence of an external bias voltage and electronic currents on
the formation and breaking of chemical bonds is a topic of increasing
importance with the continued down-scaling of electronic components.
This is especially accentuated in the limit of single-molecule devices.\cite{GaRaNi.08}

A substantial current may flow through a single bond and its effect on the
stability and impact on transport is crucial.
The phenomenon of random two-level conductance fluctuations (TLF) is generally
observed in a wide range of simple atomic and molecular
contacts.\cite{AgRoVi.93,DoMaKe.01,WaFuKi.03,ThDjOt.06,SpKrBe.10} It is often
possible to relate these to
changes in the bonding configuration driven by the current. Clearly, controlled
and reversible switching between well-defined conductance
states is a useful function.\cite{MoLi.10} Over the years many examples of
atomic\cite{EiLuRu.91,StCe.04,SpKrBe.10}
and
molecule-based\cite{StReHo.98.CouplingofVibrational,ChKaKi.06,HeMeGa.06,LiReMe.07,DaHeGo.08,HaMaCs.08,TrHuMo.09,
KuKaOk.09,MoReGr.10,OhViKe.10,BrNeTo.11,HuZhFe.11,KuShOk.12}
switches have been demonstrated. However, the understanding of how the
nonequilibrium
electronic structure impact chemical bonding and conformational changes still
pose many open questions.
First-principles calculations and comparisons with well-characterized,
time-resolved experiments can shed light on these issues.

Nonequilibrium dynamics of \buck-systems has been under intense
study\cite{PaPaLi.00,DaHeGo.08,ScFrGa.08.ResonantElectronHeating,NeKrBe.11}.
Here we focus on recently reported
time-resolved measurements of single \buck-contacts
with a scanning tunneling microscope (STM),\cite{NeKrBe.11}
which showed that TLF occur in a narrow transition regime between
tunneling and contact to \buck.
The advantage of STM is the possibility
to identify the orientation of individual \buck-molecules\cite{LaElGr.08, NeKrLi.08}
before and after controllable formation of the
tip-molecule contact.\cite{NeKrLi.07}
Moreover, the role of detailed electrode bonding geometry
\cite{ScFrBr.09,ScFrAr.11}
and contact point on the junction conductance has been
clarified.\cite{ScDaGo.11}

More specifically, the experiment revealed the following interesting properties:
(i) In the tunneling regime $dI/dV$ spectroscopy shows that transport is
dominated by the
lowest unoccupied molecular orbital (LUMO) (seen as a resonance centered at a
positive sample voltage of $\sim0.4$ V),
while (ii) in contact the $I-V$ curve is close to linear in the voltage range
$[-0.4;+0.4]$~V, suggesting a relatively
symmetric coupling of the LUMO resonance to the two electrodes. Intriguingly,
(iii) the TLF was {\em only} observed at positive sample voltage
around contact formation.
These findings were discussed in Ref.~\onlinecite{NeKrBe.11} solely on the
basis of $dI/dV$ spectra in the tunnel regime. Essentially, only the spectral properties
of the molecular adsorbate in equilibrium with the substrate were considered.
Here we present a different view on the experimental findings
based on our demonstration of a remarkably strong bias-dependent electronic coupling to
the center-of-mass (CM) motion of the \buck\ at the point when a bond is being
formed between \buck\ and the apex atom of the STM tip. From first-principles
calculations we obtain a detailed description of the \buck-junction geometry as well
as the molecular LUMO resonance near the Fermi level. This allows us to construct a model for the TLF,
which provides an explanation for the experimental findings.
Our results demonstrate that the full nonequilibrium electronic structure needs to be
accounted for to understand the observed TLF.

Our paper is organized as follows. In \Secref{sec:method} we describe the first-principles method and our setup
of the \buck-contact system. In \Secref{sec:contact-formation} we then describe the results obtained without fitting parameters for the contact
formation between STM-tip and \buck\ in equilibrium. Here we identify the formation of the chemical bond between the
molecule and the tip apex atom.
This is followed by our study of nonequilibrium effects and a discussion of the identified polarity-dependent
strong coupling between the \buck\ CM motion and voltage drop (\Secref{sec:voltage-drop}). From these first-principles
calculations we extract in \Secref{sec:heating} parameters for a simple single-resonance model, most importantly the bias-dependent
electron-vibration coupling to the CM motion. Together with a few additional parameters the model is used to calculate the
TLF behavior, which can be compared to the experiment. Before concluding we discuss how the nonequilibrium
forces modify the energy landscape for the CM-motion (\Secref{sec:current-induced-force}).

\begin{figure}
\begin{center}
\includegraphics[scale=0.42]{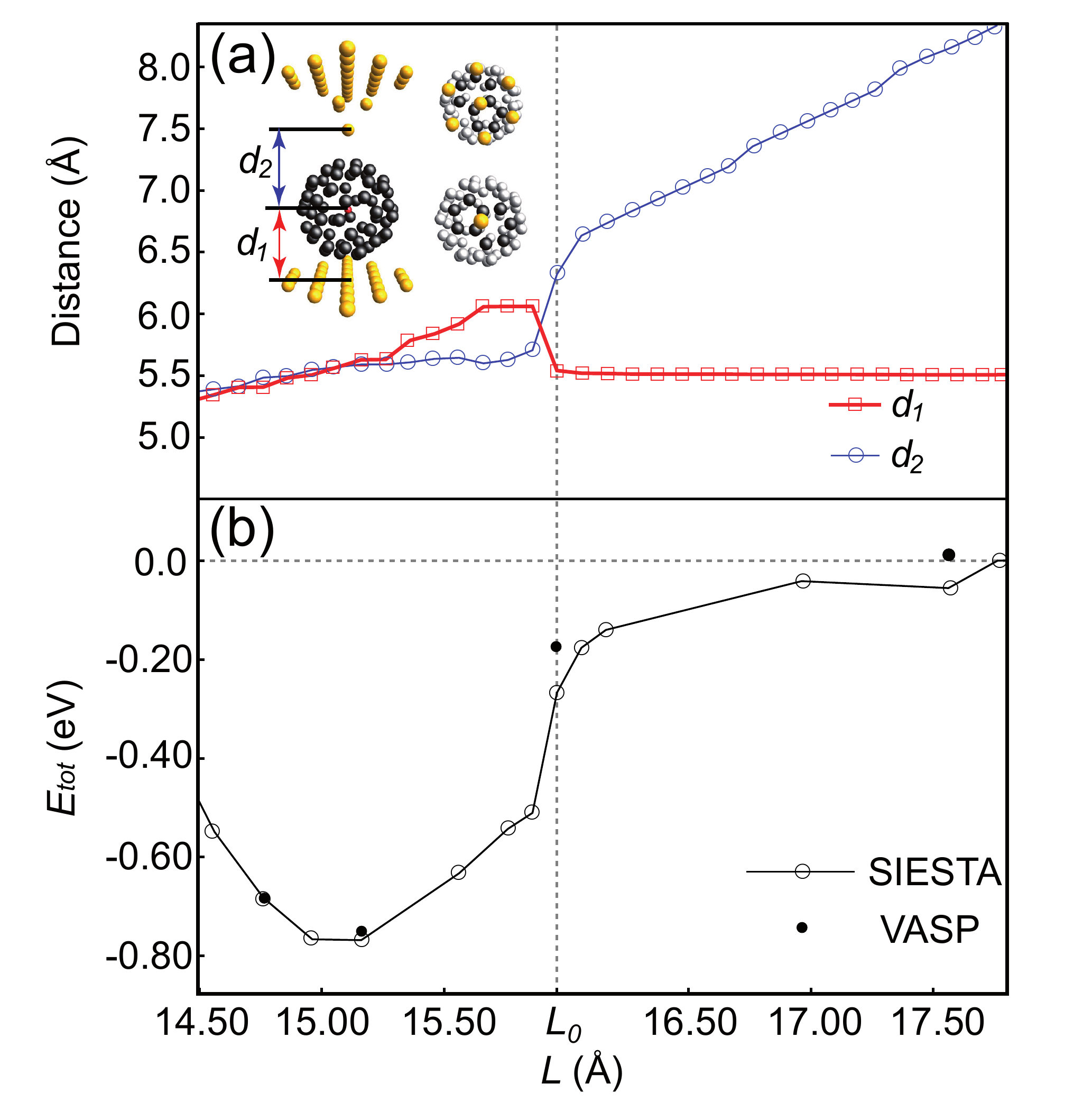}
\end{center}
\caption{(color online) (a) Relaxed bond lengths, and (b) corresponding total
energy
vs electrode separation $L$. The total energy is determined with respect to the
initial configuration.
The bond length $d_1$ ($d_2$) between C$_{60}$ center-of-mass (CM) and surface (tip) and the
junction geometry are defined in the inset
in (a) along with the orientation of \buck\ with respect to the underlying Cu
surface (top right inset) and the tip apex atom (bottom right inset).
The \buck\ is oriented such that the tip apex atom approaches the 5:6 bond.
}
\label{fig:1}
\end{figure}

\section{Method and setup}
\label{sec:method}
To study the contact formation and TLF we employ the
\textsc{Siesta}\cite{SoArGa.02}
density functional theory (DFT) method, and its extension to finite bias
using nonequilibrium Green's functions (DFT-NEGF) in the \textsc{TranSiesta}
scheme.\cite{BrMoOr.02} The generalized gradient approximation (GGA-PBE) is applied for
exchange and correlation (xc).\cite{PeBuEr.96}

The \buck\ junction geometry is modelled as shown in the inset to \Figref{fig:1}.
The periodic supercell used in the DFT calculations contains one
\buck\ molecule supported on top of seven fixed Cu(111) layers
(27 Cu atoms per layer)
with a pyramid-shaped Cu tip mounted on the bottom layer.
To accurately describe the Cu surface and the chemical bonding with \buck,
an optimized diffuse basis set was applied for Cu surface layers and the tip.\cite{optbasis}
The counterpoise correction\cite{boysbernardi.1970} for the basis set superposition errors (BSSE)
was applied to the total energy calculations, which was checked against
complementary calculations with the \textsc{Vasp} \cite{KrFu.96} plane wave
code as shown in \Figref{fig:1}(b).

The $\Gamma$-point approximation was employed for Brillouin
zone integrations in the electronic structure calculation, while
the transmission function was sampled over $3\times 3$ $\textbf{k}$-points
in the 2D Brillouin zone parallel to the electrode surfaces. The
residual atomic forces were lower than 0.02~eV/{\AA} for the atoms
that were relaxed. The \buck\ CM force constant was calculated from
DFT total energies corresponding to configurations where the \buck\
CM was rigidly displaced, up to $0.6$~\AA~from its equilibrium
position.

\section{Contact formation}
\label{sec:contact-formation}
We first focus on the bond-formation point at zero bias, and consider the
approach of the STM tip towards a 5:6 \buck-bond, {\em i.e.}, a bond between a pentagon and a hexagon.
We note that the fluctuations were observed for this orientation in the experiments,\cite{NeKrBe.11}
and that no molecular rotations occur during contact formation in either the experiments\cite{NeKrLi.08}
or in our structure optimizations.

We optimize the junction geometry by stepwise reducing the size of the DFT-supercell in the direction
perpendicular
to the surface, while relaxing the \buck\ and tip atoms.
\Figref{fig:1}(a) shows the relaxed bond lengths $d_1$ and $d_2$, between
the \buck\ center-of-mass (CM)
and the surface and the tip atoms, respectively, as a function of electrode
separation $L$.
Around a characteristic separation $L_0 = 15.96$~{\AA}, the distance $d_2$
decreases rapidly while $d_1$ increases dramatically as the cell shrinks.
This signals the onset of a chemical bond formation between the STM tip and the
\buck\ molecule.
This tip-\buck\ attraction lowers the total
energy of the system as witnessed by the binding energy curve in \Figref{fig:1}(b).

\begin{figure}
\begin{center}
\includegraphics[scale=0.35]{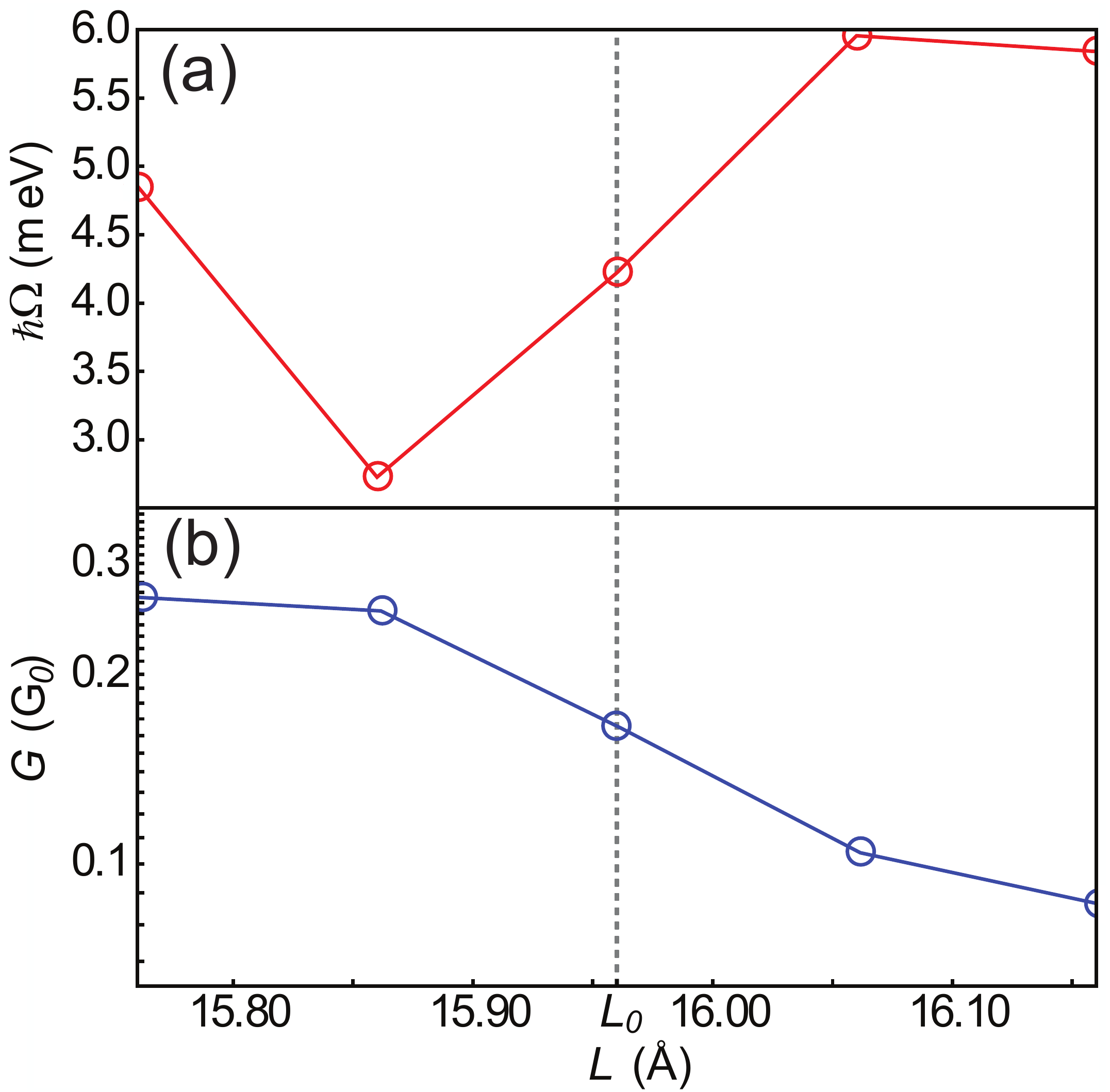}
\end{center}
\caption{(color online) (a) Vibrational energy $\hbar\Omega$ of the C$_{60}$ center-of-mass (CM)
motion between the electrodes, and (b) zero-bias
conductance $G$ of the junction vs electrode separation $L$ in the transition
regime between tunneling and contact for the DFT equilibrium geometries.}
\label{fig:2}
\end{figure}

The corresponding vibrational energy $\hbar\Omega$ associated with the \buck\ CM
motion
as well as the zero-bias conductance $G=$G$_0 T(E_F)$ (conductance quantum G$_0 = 2e^2/h$)
of the junction are shown in
\Figref{fig:2} in the transition
regime between tunneling and contact for the DFT equilibrium geometries.
At $L_0$ we find three eigenchannels contributing to the total transmission $T(E_F)$ with
the values $\{0.16,0.006,0.002\}$. The first
channel dominates the transmission, because the three-fold degeneracy of the \buck\ LUMO
has been lifted.\cite{HaDuBa.10.Calculationofimages,HeRaCh.11.EngineeringNegativeDifferential}
Thus, the \buck\ symmetry is broken in the contact configuration.
One observes that the bond formation to the tip softens the \buck-vibration [\Figref{fig:2}(a)]
and increases the conductance
by roughly a factor of 2.5 [\Figref{fig:2}(b)].  We note that the calculated conductance value of
the order $G=0.2$G$_0$
agrees very well with the experimental conductance in the transition region
between tunneling and contact
where the TLF occur.\cite{NeKrBe.11} Moreover, the
calculated vibrational energies agree
with a recent theoretical study of the  \buck\ CM-motion on the Au(111)
surface.\cite{HaArTs.12}

According to our equilibrium DFT calculations we could not identify two
well-defined stable configurations (for any fixed electrode separation)
which could explain the existence of two different conductance states. Instead we observe a
shallow energy landscape around the point of contact formation
indicating that \buck\ is rather free to move between the electrodes ({\em e.g.}, the
softening of the \buck\ CM mode).
We therefore speculate that a small barrier of the order of 10~meV, separating two distinct
configurations, could be masked by
limited numerical accuracy or by inherent approximations in the applied
xc functional. In fact, recent theoretical studies of a somewhat simpler
system consisting of graphene on Ni(111) have shown that various xc functionals
can yield differences in the potential energies describing the carbon-metal distance
much beyond the energies relevant for our system.\cite{VaMoKe.10,WeLuMo.12}
The disregard of current-induced forces acting on the atoms could also play an important role
in the energy landscape,\cite{DzKo.11} a point we return to at the end of this paper.
Finally, we note that the actual experiments involve a complex
reconstructed surface structure which we did not take into account.
Because of these circumstances we shall therefore in our TLF-model (\Secref{sec:heating})
postulate the existence of two configurations in the contact region separated by a small barrier (on the
order of DFT-accuracy), and instead focus our attention on the electron-CM vibration
coupling and the resulting current-induced heating,
which can explain the observed polarity-dependent TLF.

\begin{figure}
\begin{center}
\includegraphics[scale=0.29]{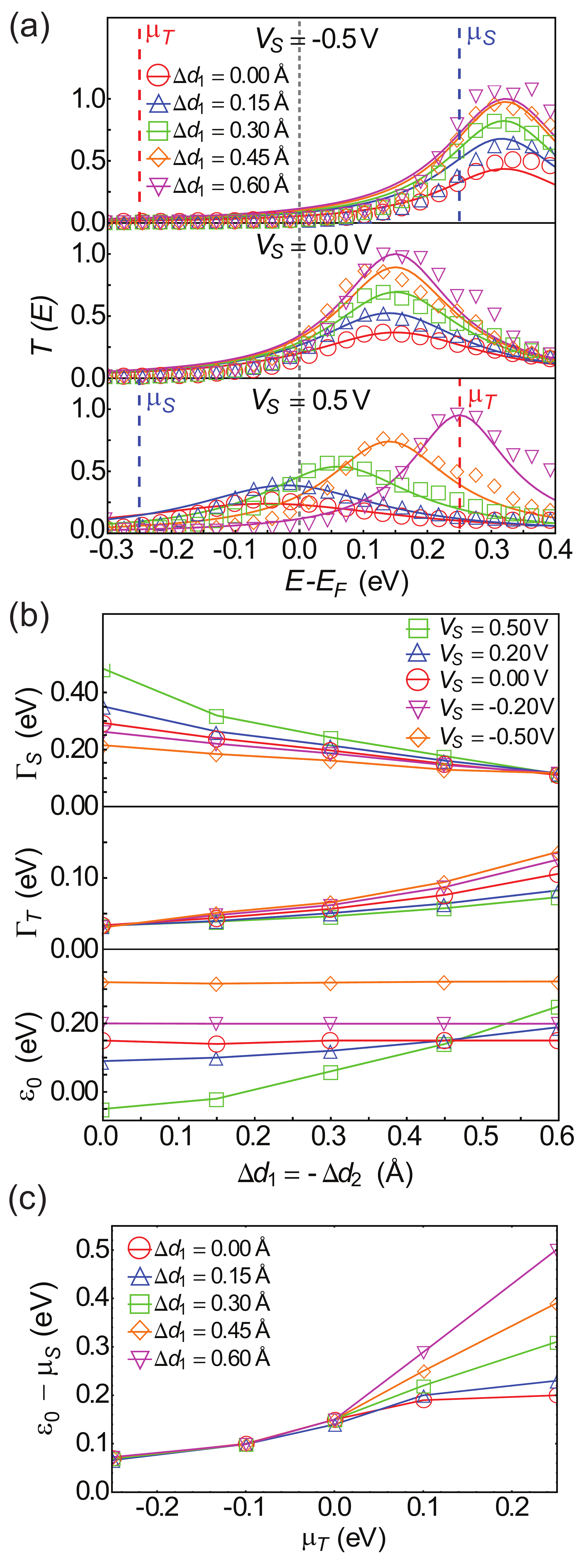}
\end{center}
\caption{(color online) (a) Transmission spectra $T(E,V_S)$ for different sample bias voltage $V_S$ as
the
C$_{60}$-surface bond length is
increased by $\Delta d_1$. Data points correspond to DFT-NEGF
simulations and solid lines are fits to the Breit-Wigner formula [\Eqref{eq:Breit-Wigner}] for the
transmission through a single molecular orbital. (b) Fitted LUMO level position
$\epsilon_0$ and coupling functions $\Gamma_S$ and $\Gamma_T$ due to the
surface and tip coupling, respectively, and (c) voltage drop across the
\buck-surface interface as a function of $\mu_T$ and $\Delta d_1$. The solid
lines in (b-c) are guides to the eye.}
\label{fig:3}
\end{figure}

As the electrode separation $L_0$ is characteristic for the point of tip-\buck\
bond formation, we take
this configuration as the starting point for an exploration of how the
nonequilibrium electronic structure and electron transport
depend on \buck\ motion.
\Figref{fig:3}(a) shows the transmission spectra (with a prominent LUMO
resonance) for several positions $\Delta d_1$ of \buck\
between the electrodes under three different applied sample voltages $V_S$.
In each situation the transmission function is approximately given by
a Breit-Wigner function\cite{Datta1995,Novaes.2011}
\begin{equation}
T(E,V_S)\approx 2\pi\frac{\Gamma_{T}\Gamma_{S}}{\Gamma_{T} + \Gamma_{S}}[\rho_T(E,V_S)+\rho_S(E,V_S)],
\label{eq:Breit-Wigner}
\end{equation}
where
\begin{equation}
\rho_{T(S)}(E,V_S)=\frac{1}{2\pi}\frac{\Gamma_{T(S)}}{[E-\varepsilon_{0}(V_S)]^{
2}+[(\Gamma_{T}+\Gamma_{S})/2]^{2}}
\end{equation}
is the partial density of states of the LUMO resonance, positioned
at $\varepsilon_0(V_S)$, due to the coupling $\Gamma_{T(S)}$ to the tip (sample)
electrode (neglecting energy dependence in $\Gamma_{T(S)}$).
We take the equilibrium Fermi energy $\varepsilon_F=0$ as the energy reference and define the tip and surface chemical potentials as $\mu_T=V_S/2$ and
$\mu_S=-V_S/2$, respectively.
The resonance parameters $\{\varepsilon_0,\Gamma_{T},\Gamma_{S}\}$ are readily
fitted to the DFT-NEGF calculations as a function of \buck-position and voltage,
as shown in \Figref{fig:3}(b).

\section{Voltage drop}
\label{sec:voltage-drop}
Remarkably, the nonequilibrium electronic structure reveals a strong variation
of $\varepsilon_0$ with \buck-position for positive sample voltages. This is a
central finding of this work and below we shall show that it can explain the
strong polarity dependence of the TLF seen in the experiments.
In \Figref{fig:3}(c) we illustrate this by plotting the change in $\varepsilon_0$
relative to $\mu_S$, as a function of $\mu_T$ for the various
\buck-displacements. For $\mu_T<0$ the $\varepsilon_0$ mainly follows $\mu_S$,
while for $\mu_T>0$ a small increase
in $d_1$ and thus coupling to the tip, makes $\varepsilon_0$ follow $\mu_T$
rather than $\mu_S$, despite $\Gamma_S > \Gamma_T$.

\begin{figure}
\begin{center}
	\includegraphics[scale=0.42]{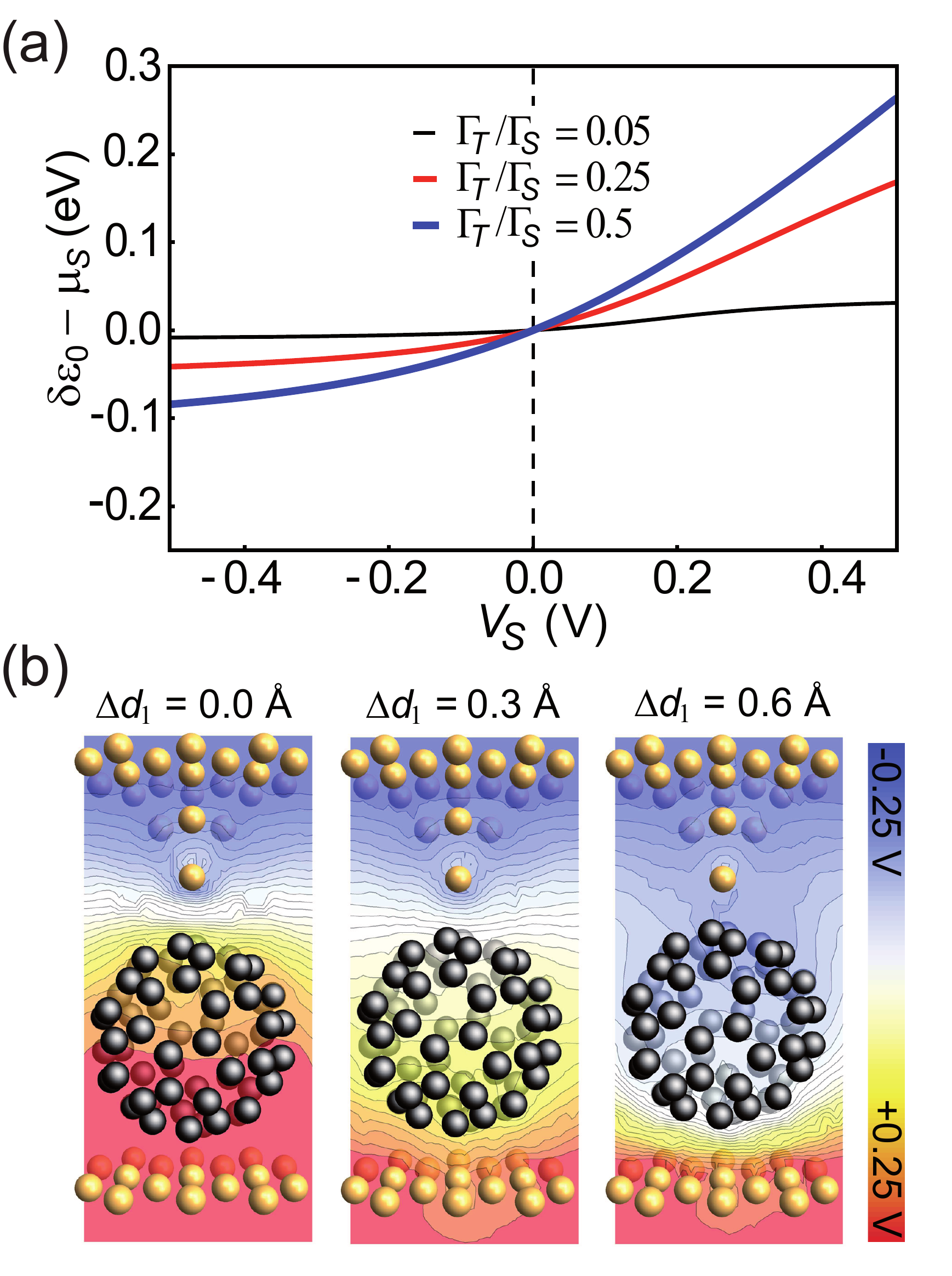}
\end{center}
\caption{(color online) (a) Simple model calculation of the change in resonance position assuming charge-neutrality of the resonance at finite bias. We use $\varepsilon_0=0.1$~eV, $\Gamma_S+\Gamma_T=0.2$~eV (fixed), and vary $\Gamma_T/\Gamma_S$ (distance to tip). The main voltage drop occurs between tip and \buck\ (resonance) for $V_S<0$ and small tip coupling, while it shifts to the \buck-substrate interface for $V_S>0$ and stronger tip coupling. (b) Voltage drop (change in DFT one-electron potential)
calculated for increasing \buck-surface bond length at a positive sample bias of 0.5~V.
The voltage difference between consecutive contour lines is 18~mV (shown in a
plane through the tip atom).}
\label{fig:fig4}
\end{figure}

The voltage dependence of $\varepsilon_0$, or equivalently the voltage profile across
the junction, can be understood roughly as a disposition of the system to
maintain a constant electron charge $Q$ in the resonance.\cite{BrKoTs.99}
In order to illustrate this we consider a simple model calculation.
Within the resonance model the LUMO charge is given by
\begin{equation}
Q(V_S)=\int_{-\infty}^{\mu_T} \rho_T(E,V_S) dE + \int_{-\infty}^{\mu_S}
\rho_S(E,V_S) dE\,.
\label{eq:charge}
\end{equation}
If we assume constant LUMO charge independent of the applied bias, i.e., $Q(V_S)=Q(0)$, we may determine the
bias-dependent change in LUMO position, $\delta\varepsilon_0(V_S)$ from \Eqref{eq:charge}.
To mimic the change in {\buck}-tip distance, $d_2$, for fixed electrode distance, $L$,
we vary $\Gamma_T/\Gamma_S$ for fixed $\Gamma_T+\Gamma_S$.  In \Figref{fig:fig4}(a)
it is seen how this simplified model reproduces the cross-over in the full DFT calculation [\Figref{fig:3}(c)]
for positive sample voltage when the contact is formed.
Thus the main voltage drop changes from being between tip and \buck\ for $V_S < 0$
to being between surface and \buck\ when $V_S>0$ and the distance to the tip is decreased ($|\delta\varepsilon_0-\mu_S| > |\delta\varepsilon_0-\mu_T|$).
From \Eqref{eq:charge} we can thus infer that in nonequilibrium there is a sensitive balance
between coupling strengths ($\Gamma_{T/S}$) and electrode chemical potentials ($\mu_{T/S}$)
that can displace the voltage drop from one interface to the other with a small relative change in coupling strengths.

The voltage drop effect can also be seen directly in the actual voltage drop landscape
(change in the one-electron
potential with respect to equilibrium) shown in \Figref{fig:fig4}(b). The voltage drop is observed to
shift
from the \buck-tip interface to the \buck-substrate interface with a small
\buck-displacement, an effect not present for $V_S < 0$ (not shown).

\section{Heating and fluctuations}
\label{sec:heating}
We next explain how the strong variation of $\varepsilon_0$ with \buck-position
for $V_S>0$
can be related to the strong polarity dependence of the TLF.
We start by assuming that the main current-dependence comes from the excitation
of \buck\ CM-motion, described by a harmonic potential with
$\hbar \Omega \approx 4$ meV [cf.~\Figref{fig:2}(a) at $L_0$].
Guided by the fact that the switching rates observed in the experiments (ms time
scale) are very slow compared to CM oscillations,
we propose that the switching involves a slow ``bottle-neck'' process, possibly
involving tunneling along the reaction coordinate (RC),
and that this process takes place when the excursion of the \buck\ ($\Delta d_1$) is
beyond some critical distance from the equilibrium position.
Inspired by the study of tunneling of a \buck\ molecule in the low-conductance regime
\cite{DaHeGo.08} we express the switching rate as
\begin{equation}
R(V_S) = r_s e^{-d_c^2/\langle \Delta d_1(V_S)^2 \rangle}=r_s e^{-\Delta/\langle E_\mathrm{CM}(V_S)
\rangle},
\label{eq:ratemodel}
\end{equation}
{\em i.e.}, as a product of the probability of \buck\ being at an excursion $\Delta
d_1=d_c$ away from equilibrium and of a rate, $r_s$, describing the slow process
along the RC. The critical distance $d_c$, or equivalently the energy barrier
$\Delta$, controls how far the \buck\ needs to move in order to facilitate
switching. The mean displacement $ \langle \Delta d_1^2 \rangle$, or equivalently the
mean oscillator energy $\langle E_\mathrm{CM}(V_S) \rangle$, are quantities which we
can calculate within our TLF-model.

\begin{figure}
\begin{center}
\includegraphics[scale=0.5]{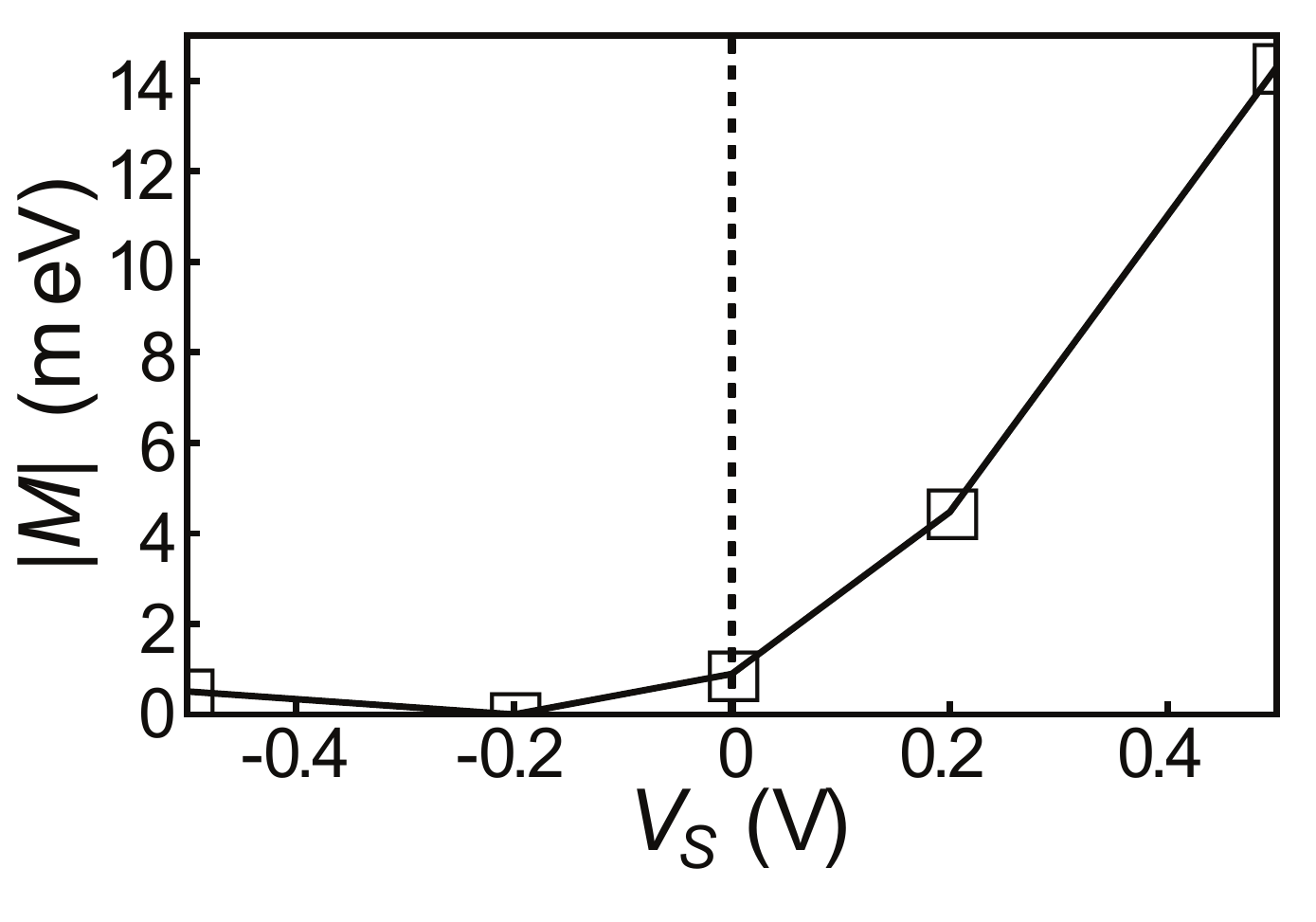}
\end{center}
\caption{(color online) Calculated electron-vibration coupling $|M|$ for the \buck\ CM motion
(black squares)
as a function of sample voltage $V_S$, based on fits to the LUMO resonance
in the transition region at the electrode separation $L_0$.}
\label{fig:5}
\end{figure}

The excitation of the \buck-CM motion by the current is determined from the
electronic coupling to this motion.
Using \Figref{fig:3}(b) we extract the electron-vibration coupling $M$ from the shift in
resonance position with \buck-displacement $d_1$ via\cite{GaPeLu.97}

\begin{equation}
M(V_S) \approx \frac{l_0}{\sqrt{2}}\, \partial_{d_1}\varepsilon_0(V_S).
\label{eq:Mep}
\end{equation}
We evaluate the slope, $\partial_{d_1}\varepsilon_0(V_S)$, around $\Delta d_1=0.3$ {\AA}, which is in the middle of the transition region [\Figref{fig:3}(b)], and note that the slope does not change significantly as we increase $\Delta d_1$.
The characteristic oscillator length is $l_0=\sqrt{\hbar/m\Omega} \approx
0.04$~{\AA} (\buck\ mass $m$), which is comparable to the size of the transition region in
\Figref{fig:2}. The extracted electron-vibration coupling, $M$, is shown in \Figref{fig:5} as
a function of sample voltage. A remarkably strong enhancement is evident for
$V_S>0$.

\begin{figure}
\begin{center}
\includegraphics[scale=0.35]{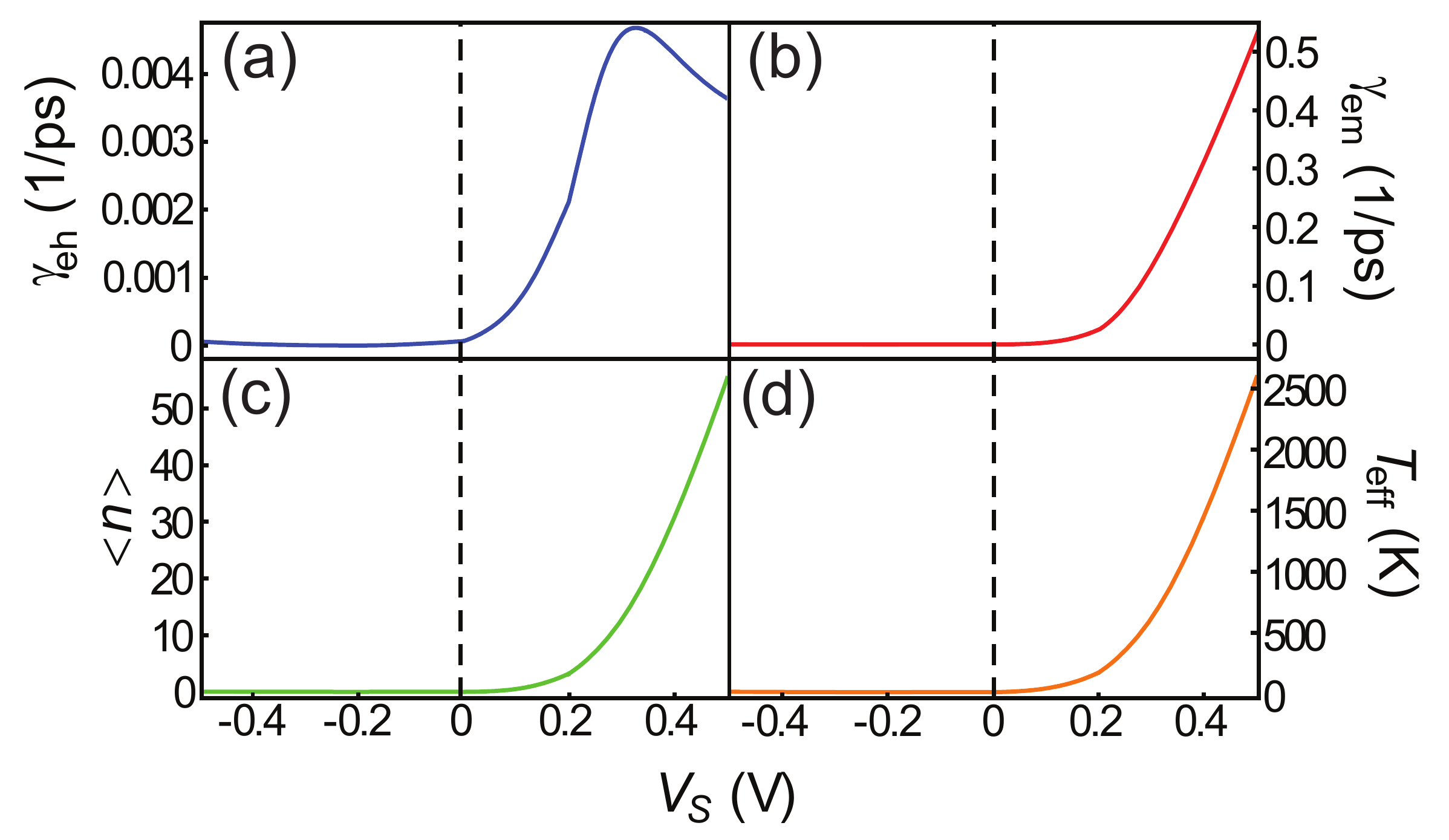}
\end{center}
\caption{(color online) Rates of (a) electron-hole pair generation ,
(b) phonon emission, and resulting (c) mean phonon occupation as a function of
sample bias for $\gamma_\mathrm{ph}\sim 0.001\Omega$. (d) Effective temperature
$T_\mathrm{eff}(V_S)$ corresponding to the  mean phonon occupation in (c).}
\label{fig:6}
\end{figure}

The excitation of the CM-motion, as seen in its mean energy $\langle E_\mathrm{CM}(V_S)
\rangle$, can be obtained from the bias-dependent
rates of phonon emission, $\gamma_{\rm em}(V_S)$, and of electron-hole pair
generation, $\gamma_{\rm eh}(V_S)$.
These rates can be determined within first order perturbation theory (Fermi's
Golden rule). Since $\hbar\Omega$ is much smaller than all other electronic
parameters, we may write
\begin{eqnarray}
{\gamma_{\rm em}}(V_S) &\approx& \frac{4\pi}{\hbar}
|M(V_S)|^{2}\theta\Big(\frac{e|V_S|}{\hbar\Omega}-1\Big)\\
&\times&\int_{-|V_S|/2}^{|V_S|/2}\, \rho_S(E,V_S) \rho_T(E,V_S)dE, \nonumber\\
\label{eq:gem}
\gamma_{\rm eh} (V_S) &\approx& 4\pi |M(V_S)|^{2} \,\Omega \\
&\times& \left[\rho(\mu_T,V_S)\rho_T(\mu_T,V_S) +
\rho(\mu_S,V_S)\rho_S(\mu_S,V_S)\right],\nonumber
\label{eq:geh}
\end{eqnarray}
where $\rho=\rho_T + \rho_S$.
From these rates we can write a rate equation for the mean
phonon occupation $\langle n (V_S)\rangle$,
\begin{equation}
\langle \dot n (V_S)\rangle = \gamma_{\rm em}(V_S)-\{\gamma_{\rm eh}(V_S)+\gamma_{\rm
ph}\}\{\langle n (V_S)\rangle - n_B\},
\end{equation}
where $\gamma_{\rm ph}$ represents the vibrational relaxation due to anharmonic
coupling to phonons in tip/substrate and
$n_B$ is the Bose-Einstein (equilibrium) phonon occupation of the considered
mode. The steady-state solution is simply
\begin{equation}
\langle n (V_S)\rangle = n_B+\frac{\gamma_{\rm em}(V_S)}{\gamma_{\rm
eh}(V_S)+\gamma_{\rm ph}}.
\end{equation}
Following Refs.~\citenum{LePe.89.Dynamicalshiftand,GaPeLu.97} one can estimate a phonon damping to the substrate of \buck-CM motion via the formula
\begin{eqnarray}
\gamma_\mathrm{ph}&=&\frac{m}{m_\mathrm{Cu}}\frac{3\pi}{2\omega_e^3}\Omega^4\approx 0.1 \Omega
\end{eqnarray}
where $m_\mathrm{Cu}$ is the mass of a substrate atom, and
$\omega_e=30$ meV a frequency characterizing the elastic response.
This damping is likely to be exaggerated compared to the experimental situation since the \buck\ is adsorbed on a
reconstructed surface with low-coordinated surface atoms and lower density of long wavelength phonons. This is a critical point for the explanation of the experimental result. We find that the best agreement is obtained for $\gamma_\mathrm{ph}\approx 0.001\Omega$. In \Figref{fig:6} we show how $\gamma_\mathrm{em}$, $\gamma_\mathrm{eh}$, and $\langle n \rangle$ varies
with the sample voltage $V_S$ along with the effective temperature defined through a Bose-Einstein distribution $\langle n(V_S)\rangle= 1/(e^{\hbar\Omega/k_BT_\mathrm{eff}(V_S)}-1)$. In all cases we see an enhancement for $V_S > 0$. If we use $\gamma_\mathrm{ph}\approx0.1 \Omega$ the mean occupation and effective temperature become a factor 100 smaller, but exhibit the same behavior as in Figs. \ref{fig:6}(c)-(d).
Finally we can calculate the oscillator energy as $\langle
E_\mathrm{CM}(V_S)\rangle=\hbar\Omega(\langle n(V_S)\rangle +1/2)$ and thus the
current-dependent rate from \Eqref{eq:ratemodel}.

\begin{figure}
\begin{center}
\includegraphics[scale=0.35]{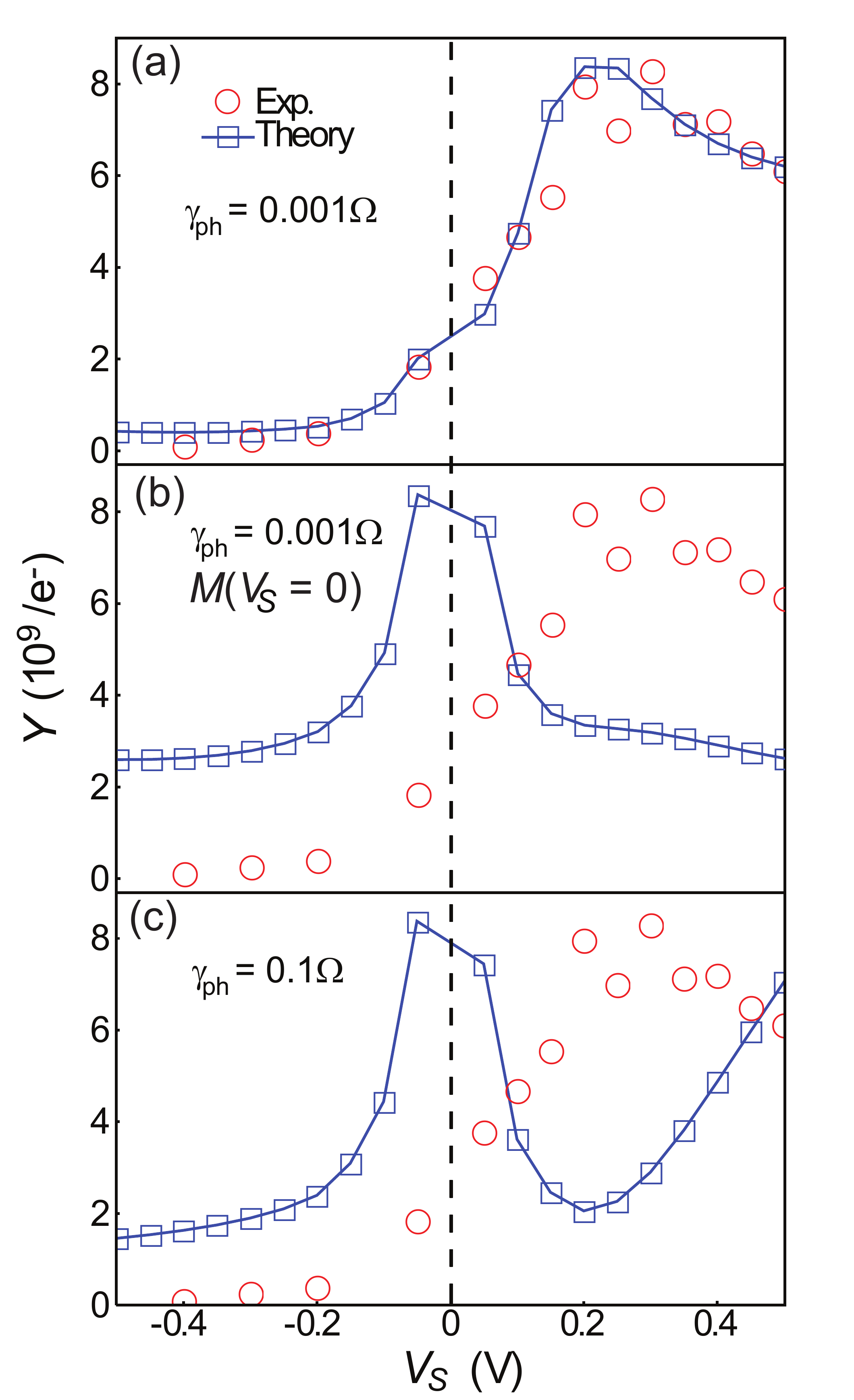}
\end{center}
\caption{(color online) (a) Calculated switching yield $Y(V_S)$ (blue squares) based on the excitation of the \buck\ CM-motion
($\hbar\Omega=4$ meV) using \Eqref{eq:ratemodel} with
$r_s = 80$ ms$^{-1}$, $T=14$ K, and $\Delta=2\hbar\Omega$. (b) Switching yield
in the case of constant electron-vibration coupling, $M(V_S)=M(V_S=0)$. (c) Same as (a) but with $\gamma_\mathrm{ph}\sim 0.1\Omega \ll \gamma_\mathrm{eh} $. For comparison also the experimental data from Ref.~\citenum{NeKrBe.11} are shown (red circles).
}
\label{fig:7}
\end{figure}

Fig. \ref{fig:7}(a) shows how the calculated switching yield $Y(V_S)$ (blue squares), defined as the switching rate per tunneling electron,
can reproduce the experimental data (red circles) if we use a barrier height of
$\Delta=2\hbar\Omega$, a ``tunnel-rate'' $r_s = 80$~ms$^{-1}$, $\gamma_\mathrm{ph}\sim 0.001\Omega$, and a background
temperature of $T=14$~K as fitting parameters.
The slightly elevated temperature, compared to the experiment performed at $T=7$
K, helps to smoothen
the onset of the switching rate at small voltages. This can be justified by
vibrational heating of other modes
and their anharmonic coupling to the CM motion of the \buck.
The relatively slow $r_s$ is consistent with a tunneling process, and the small
$\Delta$ is consistent with
the fact that we could not determine the barrier with our DFT calculations.

In \Figref{fig:7}(b) we show the calculated fluctuation rate in the case of constant zero-bias electron-vibration
coupling where only the spectral energy-dependence of the molecule are considered, cf.~the
explanation presented in Ref.~\onlinecite{NeKrBe.11} for the polarity dependence. However,
it is clear that we are only able to reproduce the experimental results
if we take the behaviour of the electron-vibration matrix element with bias into account.
These findings suggests that (i)
the strong polarity dependence of the switching
is rooted in the nonequilibrium electron-vibration coupling in the transition
region where the bond-formation between tip and \buck\ takes place,
and (ii) that the reason for the observed saturation of the switching rate per
electron is due to the steadily increasing electron-hole pair
damping with bias, \Eqref{eq:geh}, so this becomes comparable with
$\gamma_{\rm ph}$. This is an important point as illustrated in Fig. 7(c) where we show
how the switching yield using the estimate $\gamma_{\rm ph}=0.1\Omega$
grows for $V_S>0.2$ V (contrary to the experiment).

We note that the calculated current is roughly linear in voltage as in
the experiments, and thus do not contribute significantly to the polarity
dependence of the switching compared to the pronounced effect seen in \Figref{fig:5} for the electron-vibration coupling $M$.
We further note that one theoretical study\cite{SeRoGu.05} has previously reported a nonlinear, polarity-\emph{independent}
$M$ for a smaller symmetric molecular junction and only at significantly higher voltages $V>0.4$ V.

\section{Effect of current-induced force}
\label{sec:current-induced-force}
\begin{figure}
\begin{center}
\includegraphics[scale=0.35]{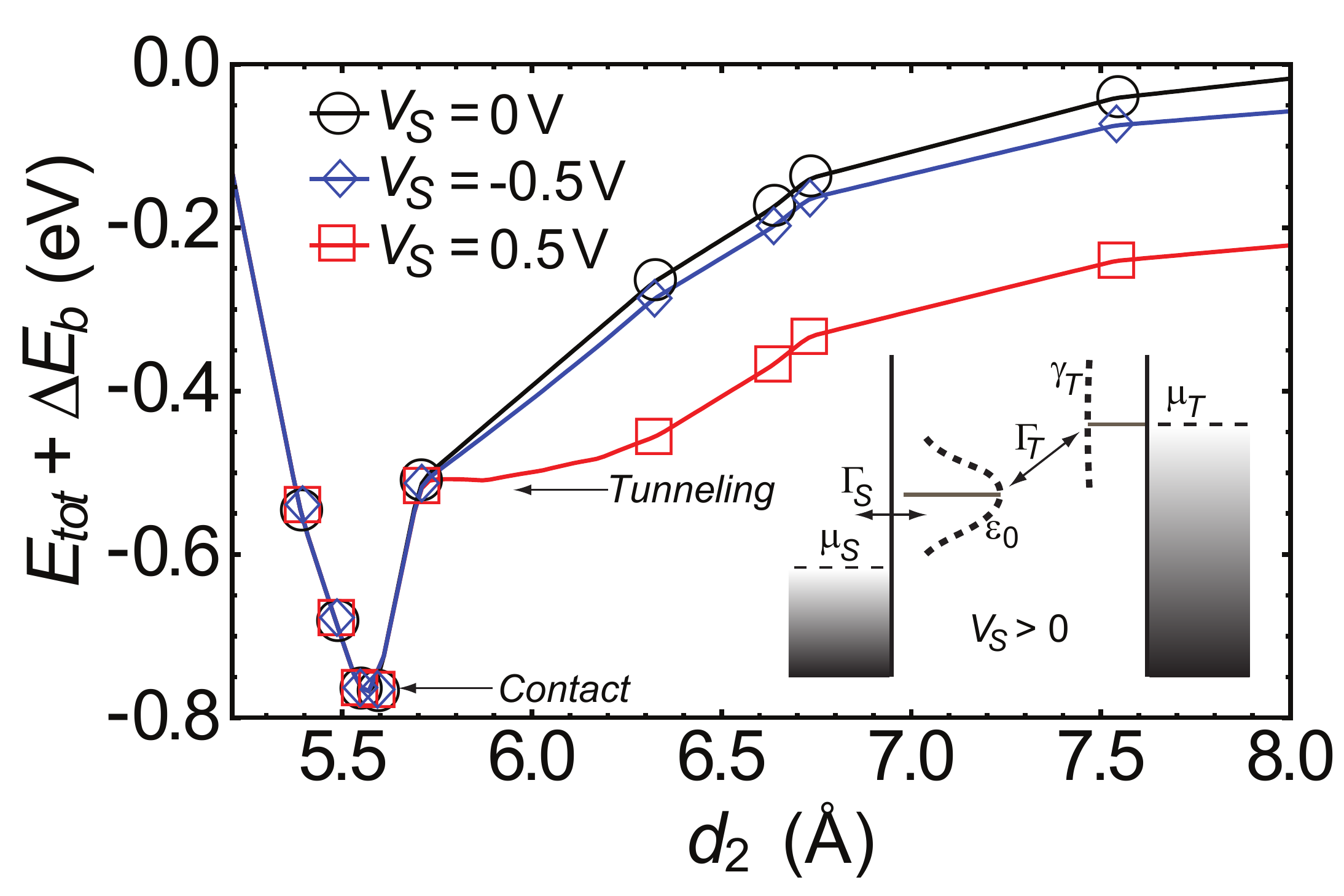}
\end{center}
\caption{(color online) Equilibrium DFT total energy from \Figref{fig:1}(b) plus an additional contribution due to a
(conservative) current-induced force. Note that energy is here plotted against \buck-tip bond
length $d_2$ instead of electrode separation. Inset: Simple model describing
a bond between the molecular resonance at $\epsilon_0$ and a wide band centered
on the tip, which follows the chemical potential $\mu_T$.
}
\label{fig:8}
\end{figure}

In this section we estimate the change in the potential energy landscape in \Figref{fig:1}(b) when
a nonequilibrium force is exerted on the \buck--tip bond during contact formation. To calculate this additional
bond force we consider the interaction between the \buck\ LUMO resonance and a wide band centered on the tip (see inset in \Figref{fig:8}).
For this system we define the following two-site Hamiltonian
\begin{equation}
\textbf{H} = \left({\begin{array}{cc}
\varepsilon_{0}(d_2,V_S) & t(d_2,V_S)\\
t(d_2,V_S) & \mu_T(V_S) \\
\end{array}}\right),
\label{eq:matsmapping}
\end{equation}
where we explicitly stress the dependence on bias and bond length $d_2$.
The interaction strength $t(d_2,V_S)$ is calculated using
$\mathrm{\Gamma}_{T}(d_2,V_S) =  2\pi\gamma_{T}|t(d_2,V_S)|^2$, where $\gamma_{T}$ is the wide band
density of states on the tip, {\em i.e.}, a constant which can be fitted to reproduce the transmission spectra in \Figref{fig:3}(a).
The bond force is calculated using the general expression\cite{ToHoSu.00}
\begin{eqnarray}
F_{b}(d_2,V_S) &=& -2\tr\left[\left(\partial_{d_2}\textbf{H}\right)\textbf{D}\right]\label{eq:overlap3}\\
 &=& -2\left[D_{11}\partial_{d_2}\varepsilon_{0}(d_2,V_S) + 2D_{12}\partial_{d_2}t(d_2,V_S) \right],\nonumber
\end{eqnarray}
where a factor of 2 is included to account for spin.
The elements, $D_{11}$ and $D_{12}$($=D_{21}$) of the density matrix $\textbf{D}$ are determined from the spectral
properties of the considered states,\cite{BrStTa.03} which can be calculated from the fits in \Figref{fig:3}(b).

Since we only consider motion along a single coordinate the current-induced force is energy conserving,
$\delta F_b(d_2,V_S) = F_b(d_2,V_S) - F_b(d_2,0)$, enabling us to calculate the change in bond energy at a given bias,
\begin{equation}
\Delta E_b(d_2,V_S) = - \int_{d_{2,i}}^{d_2} \delta F_b\left(z,V_S\right) dz.
\label{eq:newe}
\end{equation}
The integration limits are defined such that initially a contact is established, $d_{2,i} \approx 5.6$ \AA\
cf.~the equilibrium curve in \Figref{fig:8}, and then we integrate along a path where the \buck-tip contact is
gradually separated. Addition of the energy term in \Eqref{eq:newe} on top of the equilibrium total
energy in \Figref{fig:1}(b) yields the bias-corrected curves shown in \Figref{fig:8}. Astoundingly, we see that only
at positive $V_S$ a tiny barrier of the order a few meV may appear between two stable configurations
corresponding to contact and tunneling cases, respectively.
The origin of the significant lowering of the tunneling part of the binding energy curve for positive $V_S$ is related to
the asymmetry in resonance position, which yields a large
contribution from $\partial_{d_2}\varepsilon_{0}$ in \Eqref{eq:overlap3}
only at positive $V_S$. Finally, we note that the order of magnitude of the nonequilibrium barrier is in accordance with our assumption in the fluctuation calculation in \Figref{fig:7}.

\section{Conclusions}
\label{sec:conclusions}

In summary, we have presented the results of first-principles calculations which
combined with a heating model and assuming a small energy barrier can explain the experimentally observed
bias-dependent TLF observed in a \buck\ STM junction.
Our main point is that the electron-vibration
coupling can depend very strongly on the bias polarity. In this system we can trace this back
 to sensitivity of the nonequilibrium electronic structure/voltage drop
with respect to the \buck-motion just when the contact is being formed. The bias dependence of the electron-vibration coupling has so far not
been considered in most calculations of inelastic electron transport and current-induced excitations. It remains to be answered
to what extend this is important in general.
In order to model the experimental switching we had to assume a small energy barrier for the \buck-motion at the contact formation point.
Although it is likely that the small barrier is masked by inaccuracy inherent in the DFT, the finite unit-cell employed,
or numerical error, we showed that the
nonequilibrium can induce significant changes in the potential energy surface. Our estimate of the current-induced force
exerted on the \buck-tip bond did indeed indicate an energy barrier for positive sample voltage.

In the presence of a significant current a number of different excitation mechanisms can become active.
Recently, it has been discussed how current-induced forces can lead to ``run-away'' instabilities
such as bond-rupture for highly conducting systems $G\sim G_0$, and
voltages in the range involved in the present experiment\cite{DuMcTo.09,LuBrHe.10}.
TLF experiments seems to be a promising way to probe these.
The runaway effect requires the action of several vibration modes
and we have limited our discussion here to a single mode.
Our results demonstrate how the full nonequilibrium electronic structure can be of crucial
importance for the formation/breaking of chemical bonds and electron-vibration coupling in the presence of current.

\acknowledgments
We are grateful to Richard Berndt, J\"org Kr{\"o}ger, and Nicol\'as N\'eel for stimulating discussions
and comments on an early version of our manuscript. We are also thankful to Hiromu Ueba for valuable suggestions.
We also acknowledge computer resources from the DCSC.


\end{document}